\documentclass[%
 aip,
 amsmath,amssymb,
reprint%
]{revtex4-1}

\usepackage{graphicx}
\usepackage{dcolumn}
\usepackage{bm}

\usepackage[utf8]{inputenc}
\usepackage[T1]{fontenc}
\usepackage{mathptmx}

\usepackage{textcomp}
\usepackage{relsize}

\usepackage{color}
\usepackage[colorlinks=true,linkcolor=blue,citecolor=blue]{hyperref}
\usepackage[all]{hypcap}

\begin{document}

\preprint{AIP/123-QED}

\title{DC Electric Fields in Electrode-Free Glass Vapor Cell by Photoillumination}

\author{L.~Ma}
 \email{lukema@umich.edu}
\author{G.~Raithel}
 \email{graithel@umich.edu}
\affiliation{ 
Department of Physics, University of Michigan, Ann Arbor, Michigan 48109
}
\author{E. Paradis}
 \email{eparadis@emich.edu}
\affiliation{ 
Department of Physics \& Astronomy, Eastern Michigan University, Ypsilanti, Michigan 48197
}

\date{Aug 2018}%
\revised{: \today}%

\begin{abstract}

Rydberg-atom-enabled atomic vapor cell technologies show great potentials in developing devices for quantum enhanced sensors. In this paper, we demonstrate laser induced DC electric fields in an all-glass vapor cell without bulk or thin film electrodes. The spatial field distribution is mapped by Rydberg electromagnetically induced transparency spectroscopy. We explain the measured with a boundary-value electrostatic model. This work may inspire new ideas for DC electric field control in designing miniaturized atomic vapor cell devices.  Limitations and other charge effects are also discussed. 

\end{abstract}

\maketitle

\section{INTRODUCTION}

Recent efforts to coherently prepare and  precisely manipulate physical systems at a quantum level~\cite{Saffman:2019jj,Levine:2018qt,Urban:2009pl,Wade:2018lf} have led us to the dawn of a new era for applications based on laws of quantum mechanics. These applications range from fundamental quantum-based sensors\cite{Zentile:2015zg,Siddons:2009zv} and transducers\cite{Gard:2017sd,Low:2009zh} to higher level quantum networks~\cite{Cirac:1997hc} suitable for quantum communications, quantum simulation and computing\cite{Monroe:2019fk}. Among all the flourishing quantum systems, Rydberg atoms are widely regarded as a valuable building block~\cite{Saffman:2010ty,CSAdams2019} for these technologies because of their strong coupling to external electromagnetic fields~\cite{Pritchard:2010dp} and tunable interactions~\cite{Robicheaux:2018kt,Peyrot:2018xt}. 

Powered by laser cooling and coherent spectroscopy technologies~\cite{Whiting:2018gr,Sevincli:2011ey}, Rydberg atoms have demonstrated great potential in areas ranging from classical microwave~\cite{Fan:2015rq,Sedlacek:2012ue} and  THz-wave field sensing and detection~\cite{Wade:2017mr,LDownes2019,CGWage2019}  to the production of non-classical states of light~\cite{Saffman:2002hh} and readout mechanisms, such as single-photon source~\cite{Ripka:2018nr,Busche:2017oo}, photonic phase gates~\cite{Tiarks:2019mx}, etc. 

Among different working platforms, ranging from ultra cold and ultra high vacuum systems to room-temperature chip scale devices, vapor-cell-based technologies~\cite{Liew:2004yx,Daschner:2012fg,Simons:2018hm,Peyrot:2019wa} have gained significant attention over the past few decades~\cite{Phillips:2001la}. In these types of devices, Rydberg atoms are directly prepared from an ensemble of ground state atoms at moderate vapor pressures inside dielectric environments (borocilicate glasses or quartz) near room-temperature. These Rydberg atoms can be excited and proved either by cw multi-level electromagnetically-induced-transparency (EIT)\cite{Mohapatra:2007vy,Kumar:2017rr}or through pulsed nonlinear spectroscopy schemes such as four-wave-mixing~\cite{Whiting:2017ku,Ripka:2016sr}. 

These devices demonstrate amazing isolation to environmental DC electric fields outside the cell\cite{Mohapatra:2007vy,Holloway:2014nv}. These fields are generally not known but could be very large due to random electrostatic build up.  The DC field isolation is an essential prerequisite for radio frequency  (RF) and microwave sensing~\cite{Holloway:2017yo,Simons:2018go}. On the other hand, for applications such as Stark tuning of Rydberg transitions, the manipulation and control of electric fields inside the cell is necessary. By introducing metallic bulk electrodes~\cite{Grimmel:2015fx}or thin-film ITO electrodes~\cite{Barredo2013PRL} to apply fields, researchers have shown enhanced capabilities of such devices. 

In this work, we peform an experiment in which DC electric fields are produced within the dielectric enclosure without using bulk or ITO electrodes. The method presented here may enhance the functionality of the devices while still utilizing the simple structure of electrode-free dielectric vapor cells.  In addition, for applications such as RF and microwave sensing where stray DC electric fields will limit the sensitivity of the device~\cite{Holloway:2015at,Fan:2015vr}, a detailed understanding and characterization of the DC fields inside glass-only vapor cells is necessary, especially when miniaturized devices are required. In such devices, Rydberg atoms start to interact with the device enclosures~\cite{Ritter:2018gv,Whittaker:2015jx} either though quasi-static or retarded image charge interactions or by resonant Rydberg-surface plasmon coupling~\cite{Kubler:2010dn}. 

Inside Rydberg-atom-enabled vapor-cell devices, extra charges are ubiquitous~\cite{Weller:2019ys,Weller:2016wo,Abel:2011nb}.  Volume or surface charges can exist inside or on the inner surfaces of the enclosure. These charges may originate directly from Rydberg atoms through blackbody or laser induced photoionization and collision processes (Penning ionization)~\cite{Gallagher:1994qj}, and can produce volume and surface charge densities through ambipolar diffusion. Due to atomic aggregate layers on the dielectric surface, free charges can also be induced by nonuniform heating (akin to a Seebeck effect)~\cite{PhysRevLett.112.136601} and laser illumination (photoelectric effect)~\cite{osti_4712201}, leading to surface charge layers. Active control of these charging mechanisms can be used to control of the electric fields inside vapor-cell devices, and improve the Rydberg atom response in specified applications.  In this paper, we show experimental evidence that both Penning ionization and Seebeck-like effects can be suppressed and, at the same time, a photoelectric effect can be used to produce a controllable DC electric field. We also show good agreement between the measured electric field distributions and solutions of electrostatic boundary-value problems (where the geometrical effects of the enclosure boundaries are included).

\section{EXPERIMENTAL SETUP}

\begin{figure}
\begin{centering}
\includegraphics[width =3in]{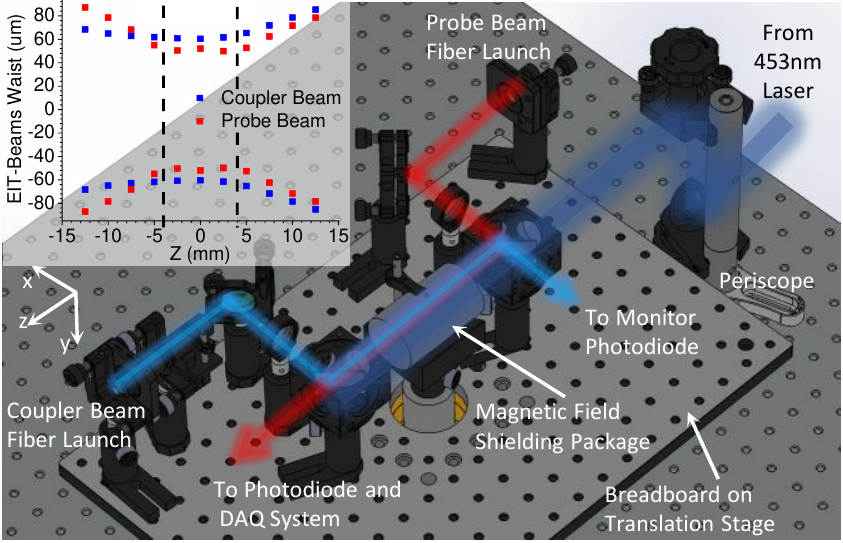}
\end{centering}
\caption{
\label{fig1}
Illustration of our experimental setup. A small part of the translation stage between the optical table and breadboard is seen in yellow. The insert shows measured optical beam waists ($1/e$ drop-off radii of the laser electric fields) of both probe and coupler beams. Two dashed lines indicate the positions of the vapor cell windows in the $z$ direction. The probe laser frequency is stabilized and $27\mathrm{MHz}$ blue-detuned from the $^{87}\mathrm{Rb}$ $5S_{1/2}, F=2$ to $5P_{3/2}, F=3$ transition.
}
\end{figure}

In order to investigate the electric field distributions inside an all-glass (borosilicate in our setup) vapor cell, DC-Stark shifts are measured through EIT resonances of a Rydberg state. The Stark shifts are mapped out as a function of position along the $x$ direction by translating the EIT channel across the vapor cell, as illustrated in Fig.~\ref{fig1}.

An  $9~\mathrm{mm}$-long and $8~\mathrm{mm}$-I.D. cylindrical glass vapor cell filled with a natural mixture of $\mathrm{Rb}$ vapor (part number VTC-11/9-10/8 from Rydberg Technologies Inc.), is placed inside a magnetic-field-shielding package. There are two $10\mathrm{mm}$ by $5\mathrm{mm}$ apertures in both end-caps of the enclosure, which allow  us to illuminate the cell walls with  a $453$-$\mathrm{nm}$ laser beam and to introduce the EIT beams. The vapor cell package is fixed to the optical table with a post that goes through a cutout in an optical breadboard on which the EIT beam optics are mounted. A translation stage connecting the breadboard and optical table is used to  translate the EIT probe region within the vapor cell relative to the cell walls.

Both the probe and coupler beams of the EIT channel are launched from single mode polarization maintaining fibers on the breadboard. The beam sizes are chosen such that we achieve the best spatial resolution under the condition that the Rayleigh length matches the vapor cell length. Measured beam sizes are plotted in the insert of Fig.~\ref{fig1}.

The $453$-$\mathrm{nm}$ photoelectric illumination beam is sent into the cell via a periscope mounted on the optical table, which allows us to adjust position of the beam relative the vapor cell. The beam size can also be adjusted via apertures (not shown in Fig.~\ref{fig1}) placed before the periscope. 

This setup provides an adjustable illumination configuration of the vapor cell. Further  the EIT channel position can be translated relative to the cell along the $x$ direction,  while maintaining a good overlap between the EIT probe and coupler beam. However, due to some warping of the vapor cell windows, slight position mismatch between probe and coupler beams is expected. This mismatch is largest when the EIT channel is brought close to the vapor cell windows edges,  where the warping occurs. This limits the EIT probing range shown in Fig.~\ref{fig2} along $x$ direction to about $6~\mathrm{mm}$, which is about $3/4$ of the inner diameter of the cell.

\section{DC Electric Fields Creation and Mapping in all glass vapor cell}

\begin{figure}
\begin{centering}
\includegraphics{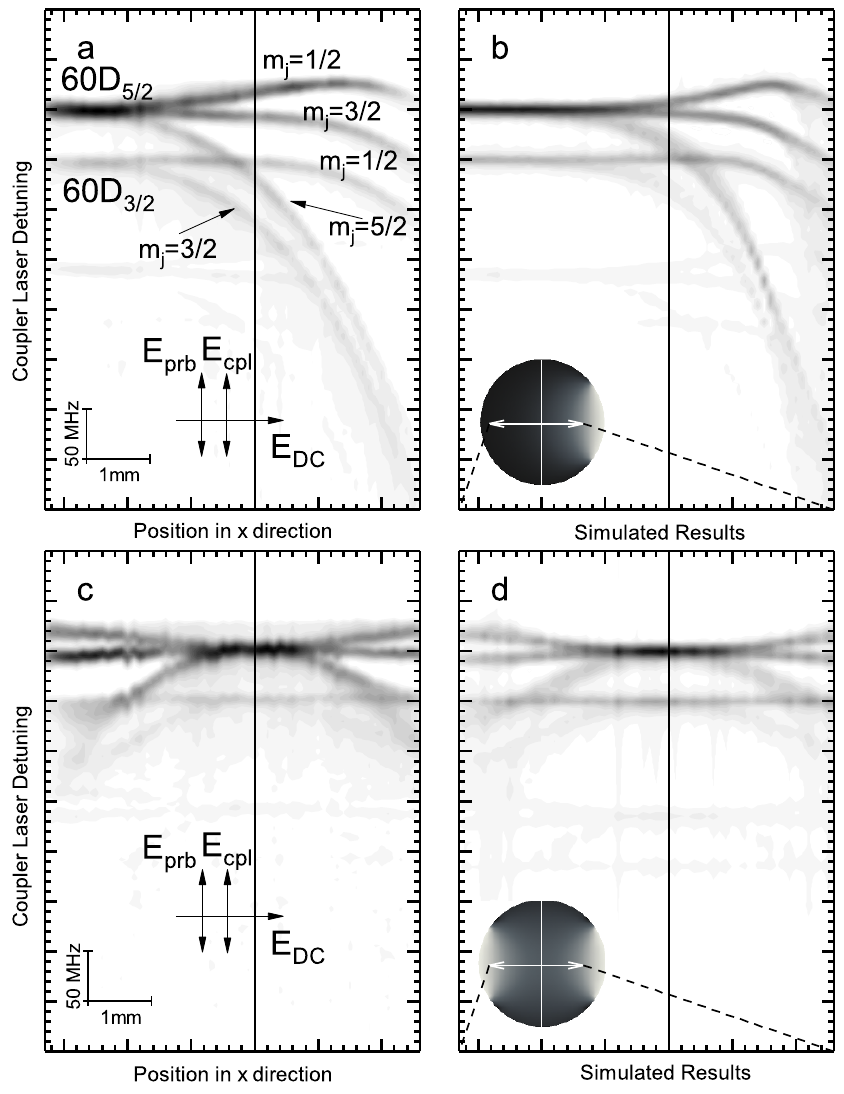}
\end{centering}
\caption{
\label{fig2}
DC Stark maps of $\mathrm{Rb}$ $60D$ measured by Rydberg EIT spectroscopy for single-sided (a) and double-sided (c) illumination of the cell walls with $453$-$\mathrm{nm}$ laser light. The probe transmission is displayed on a linear gray scale, ranging from $\sim67\%$ (white) to $\sim69\%$ (black). In our setup, the strongest EIT lines correspond to a reduction of the absorption coefficient by $\sim7\%$. DC Stark shifts of $60D_{5/2}$ and $60D_{3/2}$ sub-levels give rise to the strongest signal branches. Weak EIT signals from the intermediate $5P_{3/2}$ hyperfine sub-levels $F=2$ can also be discerned\cite{Mohapatra:2007vy}. The probe- and coupler-laser polarizations and the DC electric-field direction are indicated. Panels (b) and (d) show respective simulations, with the insets showing profiles of the electrostatic potentials on a linear gray scale ranging from 0 (black) to fitted values $V_0$ (white). The white regions on the perimeter correspond to the illuminated segments of the cylindrical cell wall. Horizontal arrows indicate the relative position between the vapor cell and probe range achieved in this work. Poor flatness on the edges of the vapor cell windows prohibits further extension of the probe range. The vertical lines indicate the geometric center of the vapor cell in $x$ direction.
}
\end{figure}

In order to achieve a high EIT signal to noise ratio the magnetic-field-shielding package is uniformly heated to $40$~\textcelsius{}. The cell walls are then illuminated by the $453$-$\mathrm{nm}$ laser beams to create photoelectric charges inside the cell. The EIT resonances of the $60D$ states of $^{87}\mathrm{Rb}$  are then mapped out as function of position of the EIT probe along $x$ direction. The Rydberg EIT resonances clearly demonstrate position-dependent DC Stark shifts. Two such DC Stark maps are shown in Fig.~\ref{fig2}~(a) and \ref{fig2}~(c), corresponding to two different photoelectric illumination conditions. Somewhat surprisingly, the Stark-split lines remain quite narrow up to considerable electric field strength, indicating a simple electric field structure that is homogeneous along the $z$ direction.

In Fig.~\ref{fig2}~(a), the $453 \mathrm{nm}$ beam is shifted to the side of the entrance aperture, so that it illuminates only one side of the cylindrical cell wall, and the illumination of the flat cell windows is negligible.  In Fig.~\ref{fig2}~(c), the $453\mathrm{nm}$ beam is centered on the aperture and the beam size is expanded to $12\mathrm{mm}$ by $12\mathrm{mm}$ before being launched from the periscope. In this case, both sides of the cylindrical cell wall are uniformly illuminated, as well as parts of the cell windows. In both cases, the top and bottom parts of the cell are not illuminated because of the 5-mm height of the optical-access slit in the cell enclosure.

In order to quantitatively understand the experimental data, we numerically simulate the density matrix by solving Master Equations based on measured optical and geometrical parameters, similar to the procedure described in Ref.\cite{Ma:2017lz}. This model includes the position-dependent Rabi frequencies in the gaussian EIT beams (see inset of Fig.~\ref{fig1}) and position-dependent Stark shifts of the atomic energy levels due to the non-uniform volume-distribution of the electric fields inside the cell. The absorption coefficient of the probe laser is calculated by using the position-averaged density matrix over all three dimensions. The simulated EIT probe beam absorption signals, shown in Fig.~\ref{fig2}~(b) and Fig.~\ref{fig2}~(d) for both illumination scenarios, are then obtained from Beer's absorption law. The only unknown parameter in this model is the electric field distribution.

\section{DC Electric Field Model} 
\label{efieldmodel}

\begin{figure}
\centering
\includegraphics{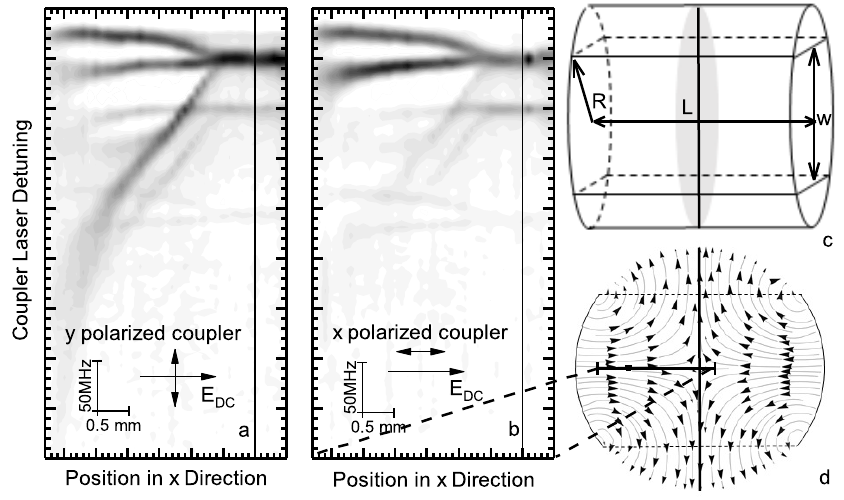}
\caption
{\label{fig3}
Measured DC Stark maps for the double-side illumination scenario when EIT coupler beam polarization is vertically aligned in the $y$ direction (a) and horizontally aligned along the $x$ direction (b). (c) Geometric model for the vapor cell used in the simulations. $R$ is the inner radius of the cell. $L$ is the inner cell length, and $w$ is height of the aperture on the magnetic field shielding end caps. In our electrostatic model, the illuminated portions of the cell walls (which are sandwiched between the two horizontal planes separated by $w$) are assigned a boundary potential value of $V_0$, the only fit parameter in our model. The electric field lines on the gray cross section are plotted in (d). Vertical lines in (a), (b) and (c) indicate the symmetry axis.  (d) Electric field lines on the gray cross section in (c). The horizontal solid line indicates the probe range shown in (a) and (b).
}
\end{figure}

The electric potential inside the cell is determined by solving the Laplace equation numerically with Dirichlet boundary conditions on the inner surfaces of  the cell walls. These boundary conditions simulate the steady state of the photoelectric processes inside the vapor cell. We assign a fixed nonzero potential $\mathrm{V_0}$ to the area that is illuminated by the $453$-nm laser, whereas the dark regions of the cell walls are set to zero potential. We also assume that there are no free space charges inside the cell. A detailed discussion follows in Sec.~\ref{discuss}. A sketch of the model for the simulations in Fig.~\ref{fig2}~(d) is depicted in Fig.~\ref{fig2}~(c) ($V_0=0.15~\mathrm{V}$). It should be noted that $\mathrm{V_0}$ is the only free parameter in this model that we use to fit the experimental data.

A second way of obtaining the potential is to perform a series expansion of the potential in a volume-charge-free cylindrical volume with end-caps~\cite{Jacksonbook}. The result for single sided illumination, used in Fig.~\ref{fig2}~(b), is shown in Eq.~\ref{eq1}; the best fit value for the surface potential is $V_0=0.33~\mathrm{V}$ in this case. We use numerical solutions of the Laplace equation and series expansions interchangeably; the results of both methods have been cross-checked for consistency.

\begin{widetext}
\begin{equation}
V(\rho,\phi,z)=\mathlarger{\sum_{m=1}^\infty}\,\mathlarger{\sum_{n=odd}^\infty}\frac{8 V_0}{mn\pi^2}\frac{I_m(n\pi\rho/L)}{I_m(n\pi R/L)}\sin\left[m\tan^{-1}\left(\frac{w}{2R}\right)\right]\sin\left(n\pi\frac{z}{L}\right)\cos(m\phi)+\mathlarger{\sum_{n=odd}^\infty}\frac{4V_0\tan^{-1}[w/(2R)]}{n\pi^2}\frac{I_0(n\pi\rho/L)}{I_0(n\pi R/L)}\sin\left(n\pi\frac{z}{L}\right)
\label{eq1}
\end{equation}
\end{widetext}

We find that the electric fields are relatively uniform in magnitude and direction within the EIT probe region, for both scenarios shown in Fig.~\ref{fig2}~(a) and Fig.~\ref{fig2}~(c). The electric-field lines on the shaded cross section in Fig.~\ref{fig3}~(c) are plotted in Fig.~\ref{fig3}~(d) for the case of double-sided illumination scenario.  The electric field exhibits the expected quadrupole characteristics in most parts of the cell, and within the EIT probe region, the field is fairly uniform. 

The direction of the DC electric field is verified by changing the polarization direction of  the coupler beam from perpendicular (as shown in Fig.~\ref{fig3}~(a)) to  parallel ( Fig.~\ref{fig3}~(b)) relative to the DC electric field. In the case of vertically polarized coupler beam, the atoms see an effective superposition of left- and right- circular polarized light relative to the $x$ quantization axis (direction of DC electric fields).This allows a much stronger $m_j=5/2$ component in the EIT spectrum (Fig.~\ref{fig3}~(a)) compare to the case in Fig.~\ref{fig3}~(b).

\section{Discussion section}
\label{discuss}

Using only one unknown parameter $V_0$, the model developed in Sec. \ref{efieldmodel} describes the electric-field distributions inside the cell relatively well. The electric potential generated by the photoelectric processes on the surfaces of a material is generally related to the energy difference between the incident photon energy (in this work, $453$-$\mathrm{nm}$ photon has a energy of $2.74 \mathrm{eV}$) and the work function of the surface. In our work, surface potentials in the range of a few hundred millivolts fit the experiment results best. Due to the complex composition of the glass materials and thin $\mathrm{Rb}$ atomic layers (aggregates) deposited on the inner surface of the glass cell, precise determination of the work function and charge affinity of such surfaces is nontrivial (see reference~\cite{Sedlacek:2016va} and citations therein) and lies outside the scope of this work.  Nevertheless, several other  (surface / free-space) charge generation mechanisms can also play a role in vapor cell Rydberg-EIT experiments, such as Penning ionization of Rydberg states, Seebeck effects , etc.

\begin{figure}
\centering
\includegraphics[width =3.37in]{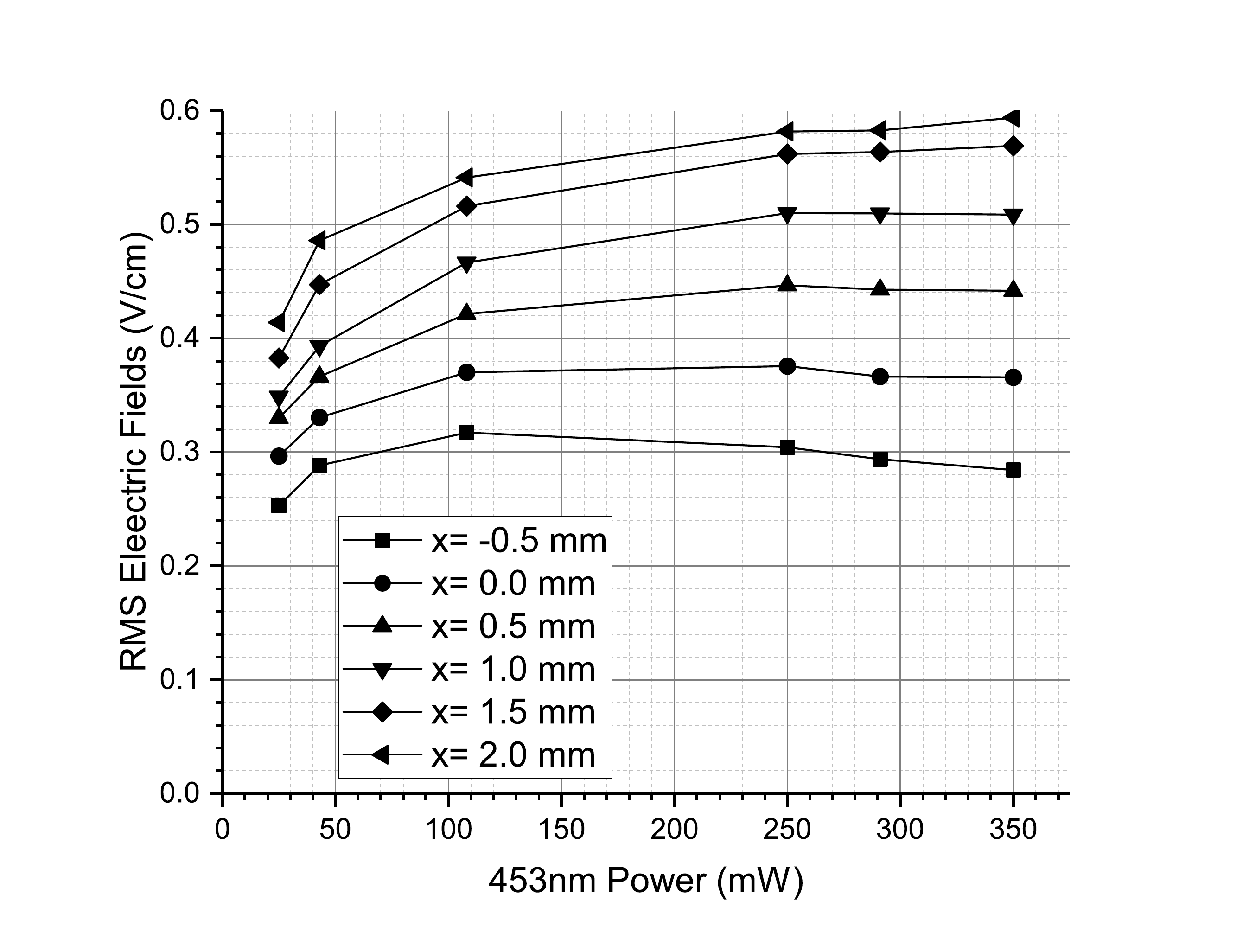}
\caption{
\label{fig4}
Measured rms value of the electric field strength as a function of the power of the $453\mathrm{nm}$ photoelectric excitation beam for the single side illuminated scenario. Different color traces represent different EIT probe position along $x$ direction. The electric field strength is obtained by matching the measured coupler beam detuning to the calculated Rydberg DC stark shifts using a large Rydberg states basis sets and direct diagonalization method. Due to the finite length of the cell, the coupler beam detuning reflects the average along the EIT propagation direction, hence the rms fields strength for fixed $y$ and $x$ position. For different probe positions in the $x$ direction, the fields demonstrate strong saturation behavior. The slight decreasing at high power range for $x$ around zero position is due to the comparable line-width of EIT transmission peak and DC stark shifts. 
}
\end{figure}

In Fig.~\ref{fig4}, the rms electric field strengths along the EIT probe propagation direction $z$ is measured as a function of incident power of the $453$-$\mathrm{nm}$ beam. Fig.~\ref{fig4} shows that for a wide range (more than one order of magnitude change) of power investigated, the rms electric fields vary only moderately, suggesting that the photoelectric charging effect is almost saturated even at lowest power used.

The other well known charge generation mechanism is Penning ionization and blackbody ionization of Rydberg atoms. This process can be adjusted by varying the power of the EIT beams. We do not observe significant changes in the DC electric field strength in the cell. (The power of both probe and coupler beams are nevertheless kept as low as possible in this work in order to achieve low power broadening and hence better spectral resolution.) These observations motivate us not to include spatial charges in our model. The results also suggest that the photoelectric effect on the cell walls is the only major source of the charges in the present work. 

We observe a decrease in electric field strength when a second $780$-$\mathrm{nm}$ beam is introduced into the system and overlapped with the $453$-$\mathrm{nm}$ beam (see supplementary materials). The second $780$-$\mathrm{nm}$ beam is tuned into resonant with $^{85}\mathrm{Rb}$ isotope  hence does not interfere with our $^{87}\mathrm{Rb}$ EIT probe beam. We believe additional charges are injected into environment caused by the direct photoionization of Rydberg states using the energy provided by the $780$-$\mathrm{nm}$ and $453$-$\mathrm{nm}$ photons. The details of the mechanism are not the focus of this work and are subject to future investigation.

It should be noted that the electric fields predicted by our model decay, as a function of position in $x$ direction, somewhat faster than fields measured in the experiment near the zero fields region for all photoelectric illumination conditions tested. This behavior is also observed for different cell temperatures. This faster decay is not a result of underestimating parameter $V_0$ and cannot be explained by taking an average of EIT beam direction along the cell length. 

We relate these observed differences to three possible causes. The first one relates to the accuracy of geometric dimensions of the vapor cells. For example, the inner surface geometry of the small nipple is not included in the model. The second relates to the underestimation of the illuminated surface area on the cell wall. In the model, this area is restricted by the aperture size of the magnetic field shielding package. Due the diffraction of $453$-$\mathrm{nm}$ beam on the edges of the aperture, the real illuminated area on the cell wall can be larger.  The third reason may relate to the non-negligible free spacial charge shielding generated by the EIT probe beam itself when it is probing the near zero fields created by the surface charges on the wall.

\section{CONCLUSION}
In this work, we have investigated the DC electric fields generated by the photoelectric process on the inner surfaces of the glass vapor cell by introducing a $453$-$\mathrm{nm}$ beam. The field distributions are mapped out using Rydberg-EIT spectroscopy. Although detailed characterization of the photoelectric effects on the cell inner surface is non-trivial, we have developed a simple phenomenological model that has reproduced the measured electric fields quite well. 

The ability to understand and manipulate the local electromagnetic fields is critical for developing miniaturized vapor cell based devices. The characterization of spurious photoelectric effects on the cell walls is important to avoid stray electric fields inside the cell. On the other hand,  the effect described in this paper can also be used to Stark-tune Rydberg transitions in vapor cells without the use of electrodes inside the cell. This functionality would allow to frequency-tune Rydberg-atom-based detectors and receivers using a laser for electric-field and Stark shift control.

{\it Acknowledgments}  This work is supported by NSF Grant PHY-1707377. We also thanks Dr. David Anderson for valuable discussions.

\bibliography{reflist}

\end{document}